\documentclass[prl,aps,twocolumn,superscriptaddress,amsmath,amssymb]{revtex4-1}

\usepackage{graphicx}

\usepackage{color}

\begin{document}

\title{Radiation Pressure on a Two-Level Atom: an Exact Analytical Approach}

\author{L. Podlecki}
\affiliation{Institut de Physique Nucl\'eaire, Atomique et de Spectroscopie, \\ CESAM, University of Liege, B\^at.~B15, Sart Tilman, Li\`ege 4000, Belgium}

\author{R. Glover}
\affiliation{Institut de Physique Nucl\'eaire, Atomique et de Spectroscopie, \\ CESAM, University of Liege, B\^at.~B15, Sart Tilman, Li\`ege 4000, Belgium}
\affiliation{Centre for Quantum Dynamics, Griffith University, Nathan, QLD 4111, Australia}

\author{J. Martin}
\affiliation{Institut de Physique Nucl\'eaire, Atomique et de Spectroscopie, \\ CESAM, University of Liege, B\^at.~B15, Sart Tilman, Li\`ege 4000, Belgium}

\author{T. Bastin}
\affiliation{Institut de Physique Nucl\'eaire, Atomique et de Spectroscopie, \\ CESAM, University of Liege, B\^at.~B15, Sart Tilman, Li\`ege 4000, Belgium}

\date{February 17, 2017}

\def\contFracOpe{%
    \operatornamewithlimits{%
        \mathchoice{
            \vcenter{\hbox{\huge $\mathcal{K}$}}%
        }{
            \vcenter{\hbox{\Large $\mathcal{K}$}}%
        }{
            \mathrm{\mathcal{K}}%
        }{
            \mathrm{\mathcal{K}}%
        }
    }
}

\begin{abstract}
    The mechanical action of light on atoms is nowadays a tool used ubiquitously in cold atom physics. In the semiclassical regime where the atomic motion is treated classically, the computation of the mean force acting on a two-level atom requires in the most general case numerical approaches. Here we show that this problem can be tackled in a pure analytical way. We provide an analytical yet simple expression of the mean force that holds in the most general case where the atom is simultaneously exposed to an arbitrary number of lasers with arbitrary intensities, wave vectors, and phases. This yields a novel tool for engineering the mechanical action of light on single atoms.
\end{abstract}

\maketitle

With the advent of lasers, the mechanical action of light has become an extraordinary tool for controlling the motion of atoms. The first evidence of this control with laser light was demonstrated in the early seventies with the deflection of an atomic beam by resonant laser radiation pressure~\cite{Sch72}. One of the most remarkable achievements was made a decade later when the first cold atomic cloud in a magneto-optical trap was observed~\cite{Chu86}. These initial experiments gave birth to the vast array of cold atom physics experiments with each more spectacular than the last. Regimes that were thought to forever remain in the realm of Gedankenexperiments became reality in labs, for example Bose-Einstein condensation~\cite{And95}.

As long as the atomic motion can be treated classically, i.e., in regimes where the atomic wavepackets are sufficiently localized in space, the resonant laser radiation acts mechanically as a force on the atomic center-of-mass. In a standard two-level approximation and modeling the laser electromagnetic field as a plane wave, the mean force $\mathbf{F}$ exerted by a single laser, averaged over its optical period, reaches shortly after establishment of the laser action a stationary regime where it takes the well-known simple expression~\cite{Cook79}
\begin{equation}
\label{F}
    \mathbf{F} = \frac{\Gamma}{2} \frac{s}{1 + s} \hbar \mathbf{k}.
\end{equation}
Here, $\Gamma$ is the rate of spontaneous decay from the upper level of the transition, $\hbar \mathbf{k}$ is the laser photon momentum, and $s = (|\Omega|^2/2)/(\delta^2 + \Gamma^2/4)$ is the saturation parameter, where $\Omega$ is the Rabi frequency and $\delta = \omega - \omega_0$ is the detuning between the laser and the atomic transition angular frequencies, $\omega$ and $\omega_0$, respectively. The maximal force the laser can exert on the atom is $(\Gamma/2) \hbar k$.

The question naturally arises of how Eq.~(\ref{F}) generalizes when several lasers of arbitrary intensities, wave vectors and phases are acting simultaneously on the atom. Surprisingly, to date no general exact analytical expression of the resulting force can be found in the scientific literature and one is often reduced to using numerical approaches~\cite{Sod97, bichr, numof}. In the low intensity regime a generalized version of Eq.~(\ref{F}) provides an approximation of the incoherent action of each laser field. Each individual laser $j$ ($j = 1, \ldots, N$ with $N$ the total number of lasers) is characterized by an individual detuning $\delta_j$, a photon momentum $\hbar \mathbf{k}_j$, a Rabi frequency $\Omega_j$, and an individual saturation parameter $s_j = (|\Omega_j|^2/2)/(\delta_j^2 + \Gamma^2/4)$. When all $s_j$ are much smaller than 1, the mean resulting force exerted incoherently by all lasers on the atom can be approximated by $\mathbf{F} = \sum_j \mathbf{F}_j$, where
\begin{equation}
\label{Fj}
    \mathbf{F}_j = \frac{\Gamma}{2} \frac{s_j}{1 + s_{\mathrm{eff}}} \hbar \mathbf{k}_j
\end{equation}
is the mean force exerted by each individual laser $j$ and $s_{\mathrm{eff}} = \sum_j s_j$~\cite{Min87}. For larger values of $s_j$, Eq.~(\ref{Fj}) loses validity as it is derived in a rate-equation approximation~\cite{Min87}. It also neglects any coherent action of the lasers. Coherence effects can lead to huge forces that vastly exceed the maximal value of $(\Gamma/2) \hbar k_j$ per laser, as is observed with the stimulated bichromatic force~\cite{Sod97}. An exact expression of the force including coherent effects and that holds at high intensity can be found in the particular case of a pair of counterpropagating lasers of same intensity~\cite{Min79}. 

In this Letter, we solve analytically the most general case and provide an expression for the force exerted by an arbitrary number of lasers with arbitrary intensities, phases,  detunings and directions acting on the same individual two-level atom. We show that the force can still be written in strict generality in the form of Eq.~(\ref{Fj}), even including coherent effects, provided a generalized definition of the saturation parameter $s_j$ is given. We thus provide a unified formalism that holds in any configuration of lasers and enables the engineering of the mechanical action of light on individual atoms.

To this end, we consider a two-level atom with levels $|e\rangle$ and $|g\rangle$ of energy $E_e$ and $E_g$, respectively ($E_e > E_g$). We denote the atomic transition angular frequency $(E_e - E_g)/\hbar$ by $\omega_{eg}$. The atom interacts with a classical electromagnetic field $\mathbf{E}(\mathbf{r}, t)$ resulting from the superposition of $N$ arbitrary plane waves: $\mathbf{E} (\mathbf{r}, t) = \sum_j \mathbf{E}_j(\mathbf{r}, t)$, with $\mathbf{E}_j(\mathbf{r}, t) = (\mathbf{E}_j/2) e^{i \left( \omega_j t - \mathbf{k}_j \cdot \mathbf{r} + \varphi_j \right)} + \textrm{c.c.}$
Here, $\omega_j$, $\mathbf{k}_j$ and $\varphi_j$ are the angular frequency, the wave vector and the phase of the $j$-th plane wave, respectively, and $\mathbf{E}_j \equiv E_j \boldsymbol{\epsilon}_j$, with $E_j > 0$ and $\boldsymbol{\epsilon}_j$ the normalized polarisation vector of the corresponding wave. The quasi-resonance condition is fulfilled for each plane wave: $\vert \delta_j \vert \ll \omega_{eg}, \forall j$, where $\delta_j = \omega_j - \omega_{eg}$ is the detuning. We define the weighted mean frequency and detuning of the plane waves by $\overline{\omega} = \sum_j \kappa_j \omega_j$ and $\overline{\delta} = \sum_j \kappa_j \delta_j$, respectively, with $\kappa_j$ a set of weighting factors ($\kappa_j \geqslant 0$ and $\sum_j \kappa_j = 1$) that can be taken a priori arbitrarily.

In the electric-dipole approximation and considering spontaneous emission in the master equation approach~\cite{Aga74}, the atomic density operator $\hat{\rho}$ obeys
\begin{equation}
    \label{masterEquation}
    \frac{d}{dt} \hat{\rho}(t) = \frac{1}{i \hbar} [\hat{H}(t),\hat{\rho}(t)] + \mathcal{D}(\hat{\rho}(t)),
\end{equation}
with $\hat{H}(t) = \hbar \omega_e |e\rangle \langle e| + \hbar \omega_g |g\rangle\langle g| - \hat{\mathbf{D}} \cdot \mathbf{E} (\mathbf{r}, t)$
and
$\mathcal{D}(\hat{\rho}) = (\Gamma/2) \left( [\hat{\sigma}_-, \hat{\rho} \hat{\sigma}_+] + [\hat{\sigma}_- \hat{\rho}, \hat{\sigma}_+] \right)$,
where $\hat{\mathbf{D}}$ is the atomic electric dipole operator, $\mathbf{r}$ is the atom position in the electric field, $\Gamma$ is the spontaneous deexcitation rate of the excited state $|e\rangle$, and $\hat{\sigma}_- \equiv |g\rangle \langle e|$ and $\hat{\sigma}_+ \equiv |e\rangle \langle g|$ are the atomic lowering and raising operators, respectively.

The hermiticity and the unit trace of the density operator make all four matrix elements $\rho_{ee}$, $\rho_{eg}$, $\rho_{ge}$, and $\rho_{gg}$, with $\rho_{kl} = \langle k | \hat{\rho} | l \rangle$ for $k,l = e,g$ dependent variables. We consider here the vector of real independent variables $\mathbf{x} = (u, v, w)^T$, with $u = \mathrm{Re}(\tilde{\rho}_{ge})$, $v = \mathrm{Im}(\tilde{\rho}_{ge})$, and $w = (\rho_{ee} - \rho_{gg})/2 = \rho_{ee} - 1/2$, where $\tilde{\rho}_{ge} = \rho_{ge} e^{-i \overline{\omega} t}$. In the rotating wave approximation (RWA) the time evolution of $\mathbf{x}$ resulting from Eq.~(\ref{masterEquation}) obeys
\begin{equation}
\label{EBO uvw n matr}
    \dot{\mathbf{x}}(t) = A(t) \mathbf{x}(t) + \mathbf{b},
\end{equation}
with $\mathbf{b} = (0,0,-\Gamma/2)^T$ and
\begin{equation}
   A(t) =
   \begin{pmatrix}
    - \Gamma/2 & \overline{\delta} & \textrm{Im} \left( \Omega (t) \right) \\
    - \overline{\delta} & - \Gamma/2 & - \textrm{Re} \left( \Omega (t) \right) \\
    - \textrm{Im} \left( \Omega (t) \right) & \textrm{Re} \left( \Omega (t) \right) & - \Gamma
    \end{pmatrix},
\end{equation}
where $\Omega(t) = \sum_j \Omega_j e^{i (\omega_j - \overline{\omega}) t}$,
with $\Omega_j$ ($j = 1, \ldots, N$) the complex Rabi frequencies 
\begin{equation}
\Omega_j = - \mathbf{D}_{ge} \cdot \mathbf{E}_j e^{i \left( - \mathbf{k}_j \cdot \mathbf{r} + \varphi_j \right)}/\hbar \equiv \Omega_{\textrm{R}, j} \, e^{i \phi_j},
\end{equation}
where $\mathbf{D}_{ge} = \langle g \vert \hat{\mathbf{D}} \vert e \rangle$, $\Omega_{\textrm{R}, j} > 0$ denotes the modulus of $\Omega_j$, and $\phi_j$ its phase. Without loss of generality, the global phases of the atomic states $|e\rangle$ and $|g\rangle$ can always be chosen so as to have one Rabi frequency real and positive. This is usually considered in all studies where a single plane wave interacts with the atom. However, with $N$ arbitrary plane waves, we cannot assume without loss of generality that all Rabi frequencies are real and their phases cannot be ignored.

Equation~(\ref{EBO uvw n matr}) constitutes the so-called Optical Bloch Equations (OBE's) adapted to the present studied case. Here, because of the time dependence of $A(t)$, the solution cannot be expressed analytically and the equation must be integrated numerically. However, within an arbitrary accuracy, we can always assume that the $N$ frequency differences $\omega_j - \overline{\omega}$ are commensurable, i.e., that all ratios $(\omega_j - \overline{\omega})/(\omega_l - \overline{\omega})$, $\forall j, l : \omega_l \ne \overline{\omega}$, are rational numbers. Within this assumption, $\Omega(t)$ and $A(t)$ are periodic in time with a period $T_c = 2 \pi / \omega_c$, where
$\omega_c = \left( \textrm{LCM} \left[ (\omega_j - \overline{\omega})^{-1} , \forall j : \omega_j \ne \overline{\omega} \right] \right)^{-1}$,
with LCM denoting the least common multiple conventionally taken as positive. It also follows that the numbers $(\omega_j - \overline{\omega})/\omega_c$, hereafter denoted by $m_j$, are integer, $\forall j$ \cite{foot1}. In the particular case where all frequencies $\omega_j$ are identical, $A(t)$ is constant in time, or equivalently periodic with an arbitrary value of $\omega_c \ne 0$, and all integer numbers $m_j$ vanish.

Within the commensurability assumption where $A(t)$ is $T_c$-periodic and given initial conditions $\mathbf{x}(t_0) = \mathbf{x}_0$, the OBE's admit the unique solution (Floquet's theorem, see, e.g., Ref.~\cite{Adr95}) 
\begin{equation}
    \mathbf{x}(t) = P_I(t) e^{R (t-t_0)} \left( \mathbf{x}_0 - \mathbf{x}_p (t_0) \right) + \mathbf{x}_p(t) ,
\end{equation}
where $R$ is a logarithm of the OBE monodromy matrix divided by $T_c$~\cite{foot2}, $P_I(t)$ is an invertible $T_c$-periodic matrix equal to $X_I(t)e^{-R(t-t_0)}$ for $t \in [t_0,t_0 + T_c]$, with $X_I(t)$ the matriciant of the OBE's~\cite{foot2}, and $\mathbf{x}_p(t)$ is an arbitrary particular solution of the OBE's. The real parts of the eigenvalues of the matrix $R$, the so-called Floquet exponents, belong to the interval $[\int_0^{T_c} q_{\textrm{min}}(t)dt/T_c, \int_0^{T_c} q_{\textrm{max}}(t)dt/T_c]$ with $q_{\textrm{min}}(t)$ and $q_{\textrm{max}}(t)$ the minimal and maximal eigenvalues of the matrix $[A(t) + A^{\dagger} (t)]/2$, respectively~\cite{Iak75}. Here, this matrix reads $\textrm{diag}(- \Gamma/2, - \Gamma/2, -\Gamma)$ and the real parts of the Floquet exponents are thus necessarily comprised between $-\Gamma$ and $-\Gamma/2$, hence strictly negative. This implies first that the matrix $e^{R (t-t_0)}$ tends to zero with a characteristic time not shorter than $\Gamma^{-1}$ and not longer than $2\Gamma^{-1}$. At long times ($t-t_0 \gg 2\Gamma^{-1}$) $\mathbf{x}(t) \simeq \mathbf{x}_p(t)$. Second, the OBE's are ensured to admit a unique $T_c$-periodic solution~\cite{Adr95}, that the particular solution $\mathbf{x}_p(t)$ can be set to. This certifies that at long times the solution of the OBE's is necessarily periodic (periodic regime).

The unique $T_c$-periodic solution of the OBE's can be expressed using the Fourier expansion
\begin{equation}
\label{FourSeriesDiffRabi}
    \mathbf{x}(t) = \sum\limits_{n = - \infty}^{+ \infty} \mathbf{x}_n e^{i n \omega_c t},
\end{equation}
with $\mathbf{x}_n \equiv (u_n,v_n,w_n)^T$ the Fourier components of $\mathbf{x}(t)$. Since $\mathbf{x}(t)$ is real, $\mathbf{x}_{-n} = \mathbf{x}_n^*$, and since it is continuous and differentiable, $\sum_n |\mathbf{x}_n|^2 < \infty$. Inserting Eq.~(\ref{FourSeriesDiffRabi}) into the OBE's yields an infinite system of equations connecting all $u_n$, $v_n$ and $w_n$ components. The system can be rearranged so as to express all $u_n$ and $v_n$ as a function of the $w_n$ components. Proceeding in this way yields, $\forall n$,
\begin{equation}
\label{unetvnavectaun}
\begin{split}
    u_n & = - i \left( \tau_n^+ \sum\limits_{j=1}^N \Omega_j w_{n - m_j} - \tau_n^- \sum\limits_{j=1}^N \Omega_j^* w_{n + m_j} \right), \\
    v_n & = - \left( \tau_n^+ \sum\limits_{j=1}^N \Omega_j w_{n - m_j} + \tau_n^- \sum\limits_{j=1}^N \Omega_j^* w_{n + m_j} \right),
\end{split}
\end{equation}
with $\tau_n^\pm = 1/[\Gamma + 2 i ( n \omega_c \pm \overline{\delta})]$ and
\begin{equation}
\label{BDSystwn}
    w_n + \sum\limits_{m \in M_0} \mathcal{W}_{n, m} w_{n + m} = \frac{- 1}{2 \left( 1 + \tilde{s} \right)} \delta_{n, 0}.
\end{equation}
Here, $\delta_{n,0}$ denotes the Kronecker symbol, $M_0$ is the set of all distinct nonzero integers $m_{lj} \equiv m_l - m_j$ ($j,l = 1, \ldots, N$),
$\mathcal{W}_{n, m} = \beta_{n, m}/(\alpha_n + \beta_{n, 0})$
($m \in \mathbb{Z}$), with $\alpha_n = \Gamma + i n \omega_c$ and $\beta_{n, m} = \sum_{j, l : m_{lj} = m} \Omega_j \Omega_l^* ( \tau_{n + m_l}^+ + \tau_{n - m_j}^- )$,
and $\tilde{s} = \sum_j \tilde{s}_j$, with
\begin{equation}
    \label{stildej}
    \tilde{s}_j = \textrm{Re} \left[ \frac{\Omega_j}{\Gamma/2 - i \delta_j} \sum\limits_{\underset{l : \delta_l = \delta_j}{l = 1}}^N \frac{\Omega_l^*}{\Gamma} \right].
\end{equation}
We have $\tau_{-n}^{\pm} = {\tau_{n}^{\mp}}^*$, $\alpha_{-n} = \alpha_n^*$, $\beta_{-n,-m} = \beta_{n,m}^*$, and $\mathcal{W}_{-n, -m} = \mathcal{W}_{n, m}^*$.

If we define the vector of all $w_n$ components for $n$ ranging from $-\infty$ to $+\infty$, $\mathbf{w} = (\ldots, w_{-1}, w_0, w_{1}, \ldots)^T$, and the infinite matrix $W$ of elements $W_{n,n'} = \sum_{m \in M_0} \mathcal{W}_{n, m} \delta_{n',n + m}$ ($n,n'$ ranging from $-\infty$ to $+\infty$), Eq.~(\ref{BDSystwn}) yields the complex inhomogeneous infinite system of equations
\begin{equation}
\label{sysinf}
(I + W) \mathbf{w} = \mathbf{c},
\end{equation}
with $I$ the infinite identity matrix and $\mathbf{c}$ the infinite vector of elements $c_n = -\delta_{n,0}/[2(1+\tilde{s})]$. $W$ is an infinite centrohermitian band-diagonal matrix with as many bands as the cardinality of $M_0$. Its main diagonal is zero. Since the OBE's admit a unique $T_c$-periodic solution, the infinite system admits a unique solution $\mathbf{w}$ with the property $\sum_n |w_n|^2 < \infty$. In these conditions and observing that $\sum_n |c_n|^2 < \infty$ and that the series $\sum_{n,n'} W_{n,n'} = \sum_{n} \sum_{m \in M_0} \mathcal{W}_{n,m}$ is absolutely convergent whatever the values of the Rabi frequencies $\Omega_j$, the detunings $\delta_j$, and the deexcitation rate $\Gamma$, the infinite system (\ref{sysinf}) can be solved via finite larger and larger truncations, whose solutions are ensured to converge in all cases to the unique sought solution $\mathbf{w}$~\cite{Rie13}. This solution is necessarily such that $w_0 \ne 0$, otherwise all other coefficients $w_n$ would solve an homogeneous system of equations and vanish, in which case the equation for $n = 0$ could not be satisfied.
This allows us to define the ratios $r_n = w_n/w_0$, $\forall n$. In particular $r_0 = 1$ and the reality condition yields $r_{-n} = r_{n}^*$. We have (Cramer's rule)
\begin{equation}
    \label{rn}
    r_n = \lim_{k \rightarrow \infty} \frac{\Delta_k^{(n)}}{\Delta_k^{(0)}},
\end{equation}
with $\Delta_k^{(n)}$ the determinant of the $I + W$ matrix truncated to the lines and columns $-k, \ldots, k$ and where the $n$-indexed column is replaced by the vector $\mathbf{c}$ correspondingly truncated. As here-above argued, the limit is ensured to exist in all cases~\cite{foot3}. Inserting $w_{m} = r_{m} w_0$ in Eq.~(\ref{BDSystwn}) for $n=0$ allows for expressing $w_0$ in the form
\begin{equation}
\label{Lastw0}
    w_0 = - \frac{1}{2} \frac{1}{1 + s_{\mathrm{eff}}} ,
\end{equation}
where $s_{\mathrm{eff}} = \sum_j s_j$, with newly defined parameters $s_j$ as
\begin{equation}
\label{FirstDefssj}
    s_j = \textrm{Re} \left[ \frac{\Omega_j}{\Gamma/2 - i \delta_j} \sum\limits_{l=1}^N \frac{\Omega_l^*}{\Gamma} r_{m_{lj}} \right].
\end{equation}

According to Ehrenfest theorem, the mean power absorbed by the atom from the $j$-th plane wave, $P_j(t) \equiv \hbar \omega_j \langle dN/dt \rangle_j(t)$, with $\langle dN/dt \rangle_j(t)$ the mean number of photons absorbed per unit of time by the atom from that wave, is given by $P_j(t) = \mathbf{E}_j(\mathbf{r}, t) \cdot d \langle \hat{\mathbf{D}} \rangle (t)/dt$. Similarly, the mean force $\mathbf{F}_j(t)$ exerted by the $j$-th plane wave on the atom reads
$\mathbf{F}_j(t) = \sum_{i = x, y, z} \langle \hat{D_{i}} \rangle \left( t \right) \nabla_{\mathbf{r}} E_{j,i}( \mathbf{r}, t )$,
with $\langle \hat{D}_i\rangle$ and $E_{j,i}( \mathbf{r}, t )$ the $i$-th components ($i = x,y,z$) of $\langle \hat{\mathbf{D}} \rangle$ and $\mathbf{E}_j(\mathbf{r}, t)$, respectively (see, e.g., Ref.~\cite{Coh80}). Of course each plane wave does not act independently of each other on the atom since the mean value of the atomic electric dipole moment is at any time determined by the atomic state $\hat{\rho}(t)$, which in turn is fully determined by the simultaneous action of all plane waves through the OBE's (\ref{EBO uvw n matr}). The total mean power absorbed from all plane waves and the net force exerted on the atom are then given by $P(t) = \sum_j P_j(t)$ and $\mathbf{F}(t) = \sum_j \mathbf{F}_j(t)$, respectively. In the RWA approximation where one neglects the fast oscillating terms, we immediately get $P_j (t) = R_j (t) \hbar \overline{\omega}$
and $\mathbf{F}_j (t) = R_j (t) \hbar \mathbf{k}_j$, with
\begin{equation}
\label{Rjt}
    R_j(t) = \textrm{Re} \left[ \Omega_j \left( v (t) + i u (t) \right) e^{i \left( \omega_j - \overline{\omega} \right) t} \right].
\end{equation}
It also follows that $\langle dN/dt \rangle_j(t) = (\overline{\omega}/\omega_j) R_j(t)$. In the quasi-resonance condition where $\omega_j \simeq \overline{\omega}$, $\forall j$, we can clearly consider $\langle dN/dt \rangle_j(t) \simeq R_j(t)$.

Within the commensurability assumption, $\omega_j - \overline{\omega} = m_j\omega_c$, $\forall j$. In the periodic regime, $u(t)$ and $v(t)$ are in addition $T_c$-periodic and thus so are $R_j(t)$, $P_j(t)$ and $\mathbf{F}_j(t)$. In this regime the Fourier components of $R_{j}(t) \equiv \sum_{n = - \infty}^{+\infty} R_{j,n} e^{i n \omega_c t}$ are easily obtained by inserting the Fourier expansion (\ref{FourSeriesDiffRabi}) into Eq.~(\ref{Rjt}). By using further Eq.~(\ref{unetvnavectaun}) and $w_n = r_n w_0$ with $w_0$ as of Eq.~(\ref{Lastw0}), we get
\begin{equation}
    R_{j,n} = \frac{\Gamma}{2} \frac{s_{j, n}}{1 + s_{\mathrm{eff}}},
\end{equation}
with $s_{j,n} = \left( \sigma_{j,n} + \sigma_{j,-n}^* \right)/2$,
where
\begin{equation}
    \sigma_{j,n} = \frac{\Omega_j}{\Gamma/2 + i \left( n \omega_c - \delta_j \right)} \sum\limits_{l = 1}^N \frac{\Omega_l^*}{\Gamma} r_{n + m_{lj}}.
\end{equation}
In particular, the temporal mean value $\overline{R}_j$ of $R_{j}(t)$ in the periodic regime is given by the Fourier component $R_{j,0}$ and observing that $s_{j,0}$ is nothing but the parameter $s_j$ of Eq.~(\ref{FirstDefssj}), we get $\overline{R}_j = (\Gamma/2)s_j/(1 + s_{\mathrm{eff}})$.
The Fourier components of the force $\mathbf{F}_j(t) \equiv \sum_{n = - \infty}^{+\infty} \mathbf{F}_{j,n} e^{i n \omega_c t}$ in the periodic regime are then given by $\mathbf{F}_{j,n} = R_{j,n} \hbar \mathbf{k}_j$~\cite{foot4} and the mean force in this regime reads consequently
\begin{equation}
    \label{finalFj}
    \overline{\mathbf{F}}_j = \frac{\Gamma}{2} \frac{s_j}{1 + s_{\mathrm{eff}}} \hbar \mathbf{k}_j,
\end{equation}
in support of our introductory claim. Here nevertheless, in contrast to the saturation parameter $s_j$ of Eq.~(\ref{Fj}), the newly defined parameter $s_j$ in Eq.~(\ref{FirstDefssj}) can be negative depending on the different phases of $\Omega_l^* r_{m_{lj}}$ with respect to $\Omega_j$. This accounts for two important physical effects. First, it can make $\overline{R}_j$ negative, in which case the atom acts as a net mean photon emitter in the $j$-th plane wave (stimulated emission) and the force exerted by that wave is directed oppositely to $\mathbf{k}_j$ (the atom is pushed in the direction opposite to the direction of the incident photons). Second, the ratio $|s_j/(1+s_{\mathrm{eff}})|$ can exceed $1$ and the force exerted by the individual laser $j$ can exceed the maximal spontaneous force $(\Gamma/2)\hbar k_j$, as is expected for coherent forces such as the stimulated bichromatic force~\cite{Sod97}.


At low intensity, i.e., for $\Omega_{\textrm{T}}^\prime \equiv \sum_j \Omega_{\textrm{R}, j}/\Gamma \ll 1$, we have $\vert \tilde{s} \vert \ll 1$ and $\sum_{n'}\vert W_{n,n'} \vert \lesssim 2 \Omega_{\textrm{T}}^{\prime 2} \ll 1$, $\forall n$. It implies that the resolvent $R_{W,-1} \equiv (I + W)^{-1}$ identifies to $I + \sum_{k=1}^{\infty} (-W)^k$ and the solution $\mathbf{w}$ of the infinite system (\ref{sysinf}) is such that $w_0 \simeq -1/[2(1+\tilde{s})] \simeq -1/2$ and $\vert w_n \vert \lesssim 2 \Omega_{\textrm{T}}^{\prime 2}$, $\forall n \ne 0$. Hence, for $m_{lj} \ne 0$, $|r_{m_{lj}}| \lesssim 4 \Omega_{\textrm{T}}^{\prime 2} \ll 1$ and it follows that $s_j \simeq \tilde{s}_j$. If all plane waves have different frequencies, the sum over $l$ in Eq.~(\ref{stildej}) only contains the single term $l = j$ and thus
\begin{equation}
\label{sjlow}
s_j \simeq \frac{\vert \Omega_j \vert^2/2}{\delta_j^2 + \Gamma^2/4}.
\end{equation}
If in contrast some plane waves have identical frequencies, coherent effects can be observed. However, if we are only interested in the incoherent action of the plane waves, an average $\langle \cdot \rangle_{\varphi}$ over all phase differences must be performed. In the low intensity regime, the statistical delta method~\cite{delMeth} yields
$\overline{R}_j^{\mathrm{inc}} \equiv \langle \overline{R}_j \rangle_{\varphi} \simeq (\Gamma/2) \langle \tilde{s}_j \rangle_{\varphi} / (1 + \langle \tilde{s} \rangle_{\varphi})$. Since the average $\langle \tilde{s}_j \rangle_{\varphi}$ is simply the $s_j$ of Eq.~(\ref{sjlow}), we get again $\overline{R}_j^{\mathrm{inc}} \simeq (\Gamma/2)s_j/(1+s_{\mathrm{eff}})$ with $s_j$ as of Eq.~(\ref{sjlow}). In all cases, the incoherent and low intensity limit of our newly defined parameter $s_j$ in Eq.~(\ref{FirstDefssj}) reduces to the standard expression (\ref{sjlow}) known to hold in this regime~\cite{Min87}.

In the particular case where all plane waves have the same frequency, $\overline{\omega} = \omega_j$, $\delta \equiv \overline{\delta} = \delta_j$ and $m_j = 0$, $\forall j$. It follows that $\Omega(t) = \sum_j \Omega_j \equiv \Omega$ and $A(t)$ is not time dependent, or equivalently is $T_c$-periodic with an arbitrary value of $T_c > 0$. The periodic regime corresponds in this case to a stationary regime where $R_j(t)$, $P_j(t)$ and $\mathbf{F}_j(t)$ are constant and determined by the zero component of their Fourier expansion. Since $m_{lj} = 0$ for any pairs of integers $l$ and $j$, the $s_j$ parameters (\ref{FirstDefssj}) simplify to $s_j = \textrm{Re} \left[ (\Omega_j/[\Gamma/2 - i \delta]) (\Omega^*/\Gamma)\right]$. For $N = 1$, this reduces to the standard expression (\ref{sjlow}) of the saturation parameter. For 2 counterpropagating plane waves, the mean net resulting force acting on the atom reads $\overline{\mathbf{F}} = (\Gamma/2) (s_1 - s_2)/(1 + s_1 + s_2) \hbar \mathbf{k}_1$. If both plane waves have identical intensity and polarization, the mean force $\overline{\mathbf{F}}$ reduces to the well known phase dependent dipole force in a stationary monochromatic wave $\overline{\mathbf{F}} = [4 \Omega_R^2 \delta \sin(\Delta \phi)]/[\Gamma^2 + 4 \delta^2 + 8 \Omega_R^2 \cos^2(\Delta \phi/2)]\hbar \mathbf{k}_1$, with $\Delta \phi = \phi_2 - \phi_1$, $\Omega_R \equiv \Omega_{R,1} = \Omega_{R,2}$~\cite{Cook79}.

Very generally, in a configuration with 2 plane waves of different frequencies, the required solution of the infinite system (\ref{sysinf}) can be obtained in a continued fraction approach (see also Ref.~\cite{Min79} for the particular case of 2 waves of identical intensity and polarization in a counterpropagating configuration). For $N = 2$ and $\omega_1 \neq \omega_2$, the commensurability assumption implies that $n_2 \kappa_1 = n_1 \kappa_2$, with $n_1$ and $n_2$ two positive coprime integers. It follows that $m_1 = \textrm{sgn}(\omega_1 - \omega_2) n_2$, $m_2 = \mathrm{sgn}(\omega_2 - \omega_1) n_1$, $m_{12} = \mathrm{sgn}(\omega_1 - \omega_2) n_s$, and $M_0 = \{ \pm n_s \}$, with $n_s = n_1 + n_2$. The infinite system~(\ref{BDSystwn}) reads in this case, $\forall n$,
\begin{equation}
\label{wnSystN2}
w_n + \mathcal{W}_{n, n_s} w_{n + n_s} + \mathcal{W}_{n, -n_s} w_{n - n_s} = \frac{- 1}{2 \left( 1 + \tilde{s} \right)} \delta_{n, 0}.
\end{equation}
The system only couples together the Fourier components $w_n$ with $n = k n_s$ ($k \in \mathbb{Z}$). All other components are totally decoupled from these former and thus vanish since they satisfy an homogeneous system. Hence, the only a priori non-vanishing ratios $r_n$ and Fourier components $R_{j,n}$ and $\mathbf{F}_{j,n}$ are for these specific values of $n$ and the periodic regime is rather characterized by the period $T_c/n_s$. For $n \neq 0$, Eq.~(\ref{wnSystN2}) implies
$w_n/w_{n-n_s} = -\mathcal{W}_{-n,n_s}^*/[1+\mathcal{W}_{n,n_s}(w_{n+n_s}/w_n)]$.
Applying recursively this relation for $n = n_s, 2 n_s, 3 n_s,  \ldots$ yields
$r_{n_s} = -\mathcal{W}_{-n_s,n_s}^*/[1+\contFracOpe_{k=1}^{\infty}(p_k/1)]$, with $p_k = -\mathcal{W}_{k n_s,n_s} \mathcal{W}_{-(k+1) n_s,n_s}^*$ and where $\contFracOpe$ stands for the continued fraction
\begin{equation}
    \contFracOpe_{k=1}^{\infty}\left(\frac{p_k}{1}\right) \equiv \cfrac{p_1}{1+\cfrac{p_2}{1 + \cfrac{p_3}{1 + \ddots}}}. 
\end{equation}
Dividing further Eq.~(\ref{wnSystN2}) for $n = k n_s$ ($k > 0$) by $w_0$ yields the recurrence relation $r_{(k+1)n_s} = - [r_{k n_s} + \mathcal{W}_{-k n_s,n_s}^* r_{(k-1)n_s}]/\mathcal{W}_{k n_s,n_s}$ that allows for computing all remaining $r_{k n_s}$ with $k > 1$, and hence along with $r_{n_s}$ all nonzero Fourier components $R_{j,k n_s}$ and $\mathbf{F}_{j,k n_s}$ ($k \geq 0$).

\begin{figure}
\includegraphics[width=0.42\textwidth]{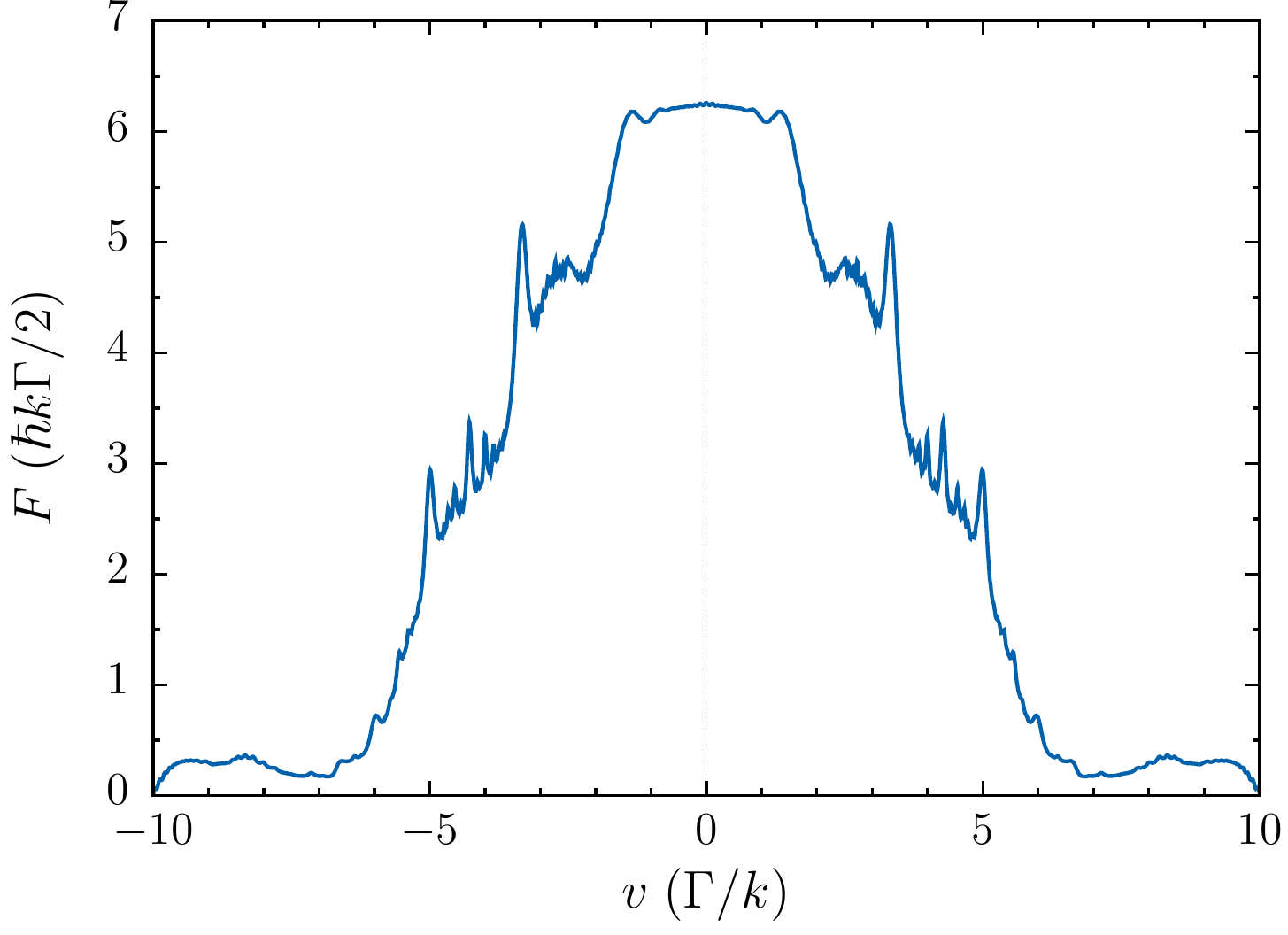}
\caption{Stimulated bichromatic force $F$ computed via our formalism for a detuning $\delta = 10 \Gamma$, a Rabi frequency of $\sqrt{3/2}\delta$ and a phase shift of $\pi/2$ for one laser.}
\label{FigBiF}
\end{figure}

Finally, we illustrate our formalism with the calculation of the stimulated bichromatic force in a standard four traveling wave configuration~\cite{Sod97}. This subtle force is complex since it relies on both coherent and high intensity effects. Except for some rough approximations, it has so far only been modeled numerically~\cite{bichr}. We considered a detuning $\delta = 10 \Gamma$, a Rabi frequency amplitude of $\sqrt{3/2} \delta$ and a phase shift of $\pi/2$ for one of the waves. We show in Fig.~\ref{FigBiF} the amplitude of the resulting bichromatic force $\sum_j \overline{\mathbf{F}}_j$ acting on a moving atom as a function of its velocity $v$. As expected, the peak value of the bichromatic force is of the order of $(2/\pi)(\delta/\Gamma)$ (in units of $\hbar k \Gamma/2$) and spans a velocity range of the order of $\delta/\Gamma$ (in units of $\Gamma/k$)~\cite{Cas03}. A direct numerical integration of the OBE's completed with a numerical average of $R_j(t)$ in Eq.~(\ref{Rjt}) in the periodic regime produces identical results.

In conclusion, we have provided a unified and exact formalism to calculate the mechanical action of a set of arbitrary plane waves acting simultaneously on a single two-level atom. We have shown that the light forces can still be written in strict generality in the form of Eq.~(\ref{Fj}), including coherent and high intensity effects, provided the generalized definition (\ref{FirstDefssj}) of the parameter $s_j$ is given. These results provide a novel tool for engineering the mechanical action of lasers on individual atoms.

T.B. acknowledges financial support of the Belgian FRS-FNRS through IISN grant 4.4512.08 and, with R.G., of the Interuniversity Attraction Poles Programme initiated by the Belgian Science Policy Office (BriX network P7/12). L.P. acknowledges an FNRS grant and the Belgian FRS-FNRS for financial support.


\begin{thebibliography}{99}
\bibitem{foot1} All nonzero $m_j$ integers are setwise coprime (the greatest common divisor of these integers is 1) but not necessarily pairwise coprime (each pair of nonzero $m_j$ are not necessarily setwise coprime).
\bibitem{foot2} The monodromy matrix $C$ of a set of equations $\dot{\mathbf{x}} = A(t) \mathbf{x} + \mathbf{b}(t)$ where $A(t)$ and $b(t)$ are $T$-periodic (comprising the case $b(t)$ constant) is the matriciant $X_I(t)$ of the set of equations evaluated at time $t = t_0 + T$: $C = X_I(t_0 + T)$. The matriciant is the unique matrix solution $X(t)$ to the matricial equation $\dot{X} = A(t) X$ subscribed to the initial conditions $X(t_0) = I$, with $I$ the identity matrix and where $X$ has the same dimension as $A(t)$. The matriciant and monodromy matrices are both invertible. A logarithm of an invertible matrix $B$ is a matrix $L$ such that $e^L = B$. Every invertible matrix admits a logarithm, which is not unique but with invariant real parts of the eigenvalues.
\bibitem{foot3} As a consequence of the linear system~(\ref{sysinf}) that connects $w_0$ to $w_{m_{lj}}$, $\forall l,j$, then to $w_{m_{lj} + m_{l'j'}}$, $\forall l',j'$, and so on, the convergence in Eq.~(\ref{rn}) is highly optimized by reordering the vector of unknowns $\mathbf{w}$ (and accordingly the lines and columns of $I + W$) so as to have $w_0$ symmetrically surrounded by $w_{\pm|m_{lj}|}$ for all distinct nonzero $m_{lj}$, then by $w_{\pm|m_{lj} + m_{l'j'}|}$ for all additional distinct $m_{lj} + m_{l'j'}$, and so on.
\bibitem{foot4} The specific links between the unknowns $w_n$ in the linear system~(\ref{sysinf})~\cite{foot3} can make the ratios $r_{n+m_{lj}}$ and hence the Fourier components $\mathbf{F}_{j,n}$ not necessarily significant for the first $n > 0$ values. In this case, the forces $\mathbf{F}_j(t)$ could be better described in the periodic regime by an almost periodic behavior with an almost period smaller than $T_c$ depending on each specific case.
\end{thebibliography}
\end{document}